\documentclass[a4paper, twocolumn, showpacs,reprint, aps,pra, superscriptaddress]{revtex4-1}

\usepackage{amsmath, amssymb, mathtools, physics}
\usepackage{hyperref}
\usepackage{times,graphicx,xcolor}
\usepackage{subfigure}
\usepackage{xspace}

\newcommand*{\cf}{\mbox{cf.}\xspace}
\newcommand*{\eg}{\mbox{e.\,g.}\xspace}
\newcommand*{\ie}{\mbox{i.\,e.}\xspace}

\newcommand*{\up}{\uparrow}
\newcommand*{\down}{\downarrow}
\newcommand*{\initial}{\mathrm{i}}
\newcommand*{\final}{\mathrm{f}}
\newcommand*{\prob}{\mathbb{P}}
\newcommand*{\equi}{\mathrm{eq}}

\begin{document}

\title{Fluctuation Relation for Qubit-Calorimetry}
\date{\today}

\author{Antti~Kupiainen}
\email{antti.kupiainen@helsinki.fi}
\author{Paolo~Muratore-Ginanneschi}
\email{paolo.muratore-ginanneschi@helsinki.fi}
\affiliation{
	University of Helsinki, Department of Mathematics and Statistics
    P.O.~Box 68 00014, Helsinki, Finland
}
\author{Jukka~Pekola}
\email{jukka.pekola@aalto.fi}
\affiliation{
	Aalto University, School of Science, 
	P.O.~Box 13500, 00076 Aalto, Finland
}
\author{Kay Schwieger}
\email{kay.schwieger@helsinki.fi}
\affiliation{
	University of Helsinki, Department of Mathematics and Statistics
    P.O.~Box 68 00014, Helsinki, Finland
}

\begin{abstract} 
	Motivated by proposed thermometry measurement on an open quantum system, we
	present a simple model of an externally driven qubit interacting with a finite
	sized, fermion environment acting as calorimeter. The derived dynamics is
	governed by a stochastic Schr\"odinger equation coupled to the temperature
	change of the calorimeter. We prove a fluctuation relation and deduce from
	it a notion of entropy production. Finally, we discuss the first and second
	law associated to the dynamics.
\end{abstract}

\maketitle


\section{Introduction}

How the thermodynamic laws of the macroscopic world transduce and impinge the
behavior of structure on the nano- and quantum-level structures, was a question
recurrently raised in the past by speculation on the fundamental bounds imposed
by physical laws on information processing and transfer
\cite{Lan61,Bek81,LLo00,LeffRex03}.

Developments of the last decade have demonstrated the experimental
feasibility of temperature measurements in nanoscale systems with sub-microsecond time resolution. For example, high
sensitivity temperature measurements have been accomplished by embedding a
superconductor-insulator-normal metal (SIN) tunnel junction into a
radio-frequency (RF) resonant circuit. The temperature of
the normal metal side of an SIN tunnel junction is then read by measuring the
reflection or transmission coefficient of the LC resonator as a function of both the
temperature and of an external current bias \cite{ScYuCl03} (see
\cite{GaViSaFaArMePe15} and the review \cite{GiHeLuSaPe06}).
SIN tunnel junction thermometers can be employed as fast calorimeters to
perform basic studies of the thermodynamics of a mesoscopic nanostructure. As a
consequence, questions which were restricted to speculation only a couple of
decades ago are now becoming feasible in properly designed experiments.

A fairly recent proposal is the calorimetric measurement of the full
distribution of the work done by exerting an external drive on a quantum two
level system (qubit) \cite{PeSoShAv13}. Let us shortly recall the setup which
makes the proposed experiment realistically possible. The qubit is implemented
using a solid state electronic circuit \cite{ClWi08}. A basic example is a
superconducting Cooper pair box (CPB) \cite{AvZoLi84,But87,MaScSh99,ClWi08} or
a transmon qubit \cite{KoYuGaHoSc07,ScHoKoScJo08}. A~CPB consists of a
superconducting electrode (island) put in contact with a superconducting
reservoir. Actual implementations use, \eg, a
superconductor-insulator-superconductor (SIS), aluminium-aluminium
oxide-aluminium $(\mathrm{Al}/\mathrm{Al}_2\mathrm{O}_{3}/\mathrm{Al})$, tunnel
junction with capacitance~$C_{j}$. Charges are driven from the reservoir to
the island by a voltage source $V$  between the reservoir and a
capacitor~$C_{g}$ connected to the island. If the energy of thermal
fluctuations $k_{\mathrm{B}}T$ and the charging (Coulomb) energy of the island
are much smaller than the superconducting gap $\Delta$ (which is about 1 K for
aluminium), all electrons in the island are paired. The experiment thus needs
to be performed at temperatures of the order of 0.1 K or below.  The effective
Coulomb energy of the island becomes $E_{C}=(2\,e)^{2}/2(C_{g}+C_{j})$.
Finally, tuning the circuit parameters so that $E_{C}\gg\,k_{\mathrm{B}}\,T$
ensures that only the ground state (no pair in the island) and the first
excited state (one pair in the island) have non-negligible probability. A~qubit
dynamics is thus effectively realized (\cite{BoViJoEsDe98,NaPaTs99}, see  also
\cite{Nak09} and refs.{} therein). 

A basic calorimeter is formed of a normal metal island with typically $10^9$
electrons on it. It is only weakly coupled to the surrounding thermal
bath (phonons) at these low temperatures. Calorimetric measurements of the
solid state qubit just described can be then performed by connecting the normal
metal electrode as a SIN junction thermometer
\cite{PeSoShAv13,GaViSaFaArMePe15} to the resonant circuit. 
The experimental setup of \cite{PeSoShAv13} envisages
a drive signal having equal free-energy at the end of the control horizon.
Energy conservation then implies that up to boundary terms the work $W$ done on
the system is equal to the heat $Q$ dissipated to the environment in the time
interval from the beginning of the driving till the end of the equilibration
period.  

In classical non-equilibrium thermodynamics, a group of relations commonly
referred to as fluctuation theorems
(\cite{EvSe94,GaCo95,Jar97,Cro97,Kur98,LeSp99}, see also
\cite{JiangQianQian,ChGa08,Sekimoto} for review) links the distribution of the
work done on a small system during a transition from an equilibrium state to
the free energy of the system at the transition end-states. Classically, the
work is a random functional of the protocol, \ie, the sequence of
non-conservative forces exerted on the system to drive the transition. From the
mathematical slant, fluctuation relations are statements about deviations in  
the distribution of work around zero. They are therefore obtained by comparing
the work distribution during the transition and a second one specified by an
appropriately defined time reversed protocol \cite{ChGa08}.

The extension of fluctuation theorems to the quantum case poses the problem
of how to define the work as a functional of the path that 
the system has followed during the transition. Intense research activity 
recently focused to determine work for closed
and open quantum systems (see \cite{Kur00,Mu03,DeRoMa04,JaWo04,AtGa09,
CaHaTa11,ChMa12,HoPa13,AlLiMaZa13,SoAvPe13}). 
For this reason the experiment proposed in \cite{PeSoShAv13} stirs interest. 
It calls for a detailed analysis of how fluctuation relations can be derived and
used to achieve a consistent non-equilibrium thermodynamic 
description of a driven qubit.
This is the scope of the present work.

\begin{figure}
	\centering
	\includegraphics[width=8.0cm]{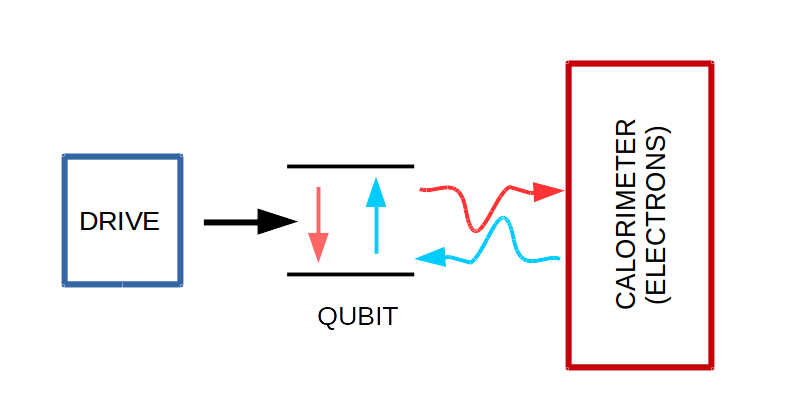}
	  \label{fig:model}
	\caption{
Pictorial representation of the theoretical model described in sec.~\ref{sec:model}.
	}
	\label{fig:Fig}
\end{figure}

The paper contains two main results. The first is to present a detailed
theoretical model for the driven qubit-calorimeter dynamics  of \cite{PeSuGa16}
(see fig.~\ref{fig:Fig}). In particular, in section~\ref{sec:model} we derive
the stochastic Schr\"odinger equation (\cf \cite{BreuerPetruccione,WiMi09})
describing the evolution of the quantum trajectories followed by a quantum
system continuously monitored by an environment, in our case embodied by the
calorimeter. In the experimental setup of \cite{PeSuGa16}, temperature
measurements are possible because the calorimeter can be thought of as a fermion
system with a number of degrees of freedom, of the order $\mathcal O(10^9)$ larger than the qubit
but \emph{finite}. In consequence, the changing
temperature effects the coupling between qubit and calorimeter. Hence, the
novelty of our model is to derive in addition to the equation for the system
evolution its coupling to the evolution law of the temperature of the
calorimeter (\cf \cite{BeBrSa15,PeSuGa16,SuKuAlNi16}).

After analyzing the measurement protocol in section~\ref{sec:path_distr}, we
derive in section~\ref{sec:fr} the fluctuation relation satisfied by our system
under what appears the natural notion of time reversal in our setup (see
\cite{PeSoShAv13}). We then use the fluctuation relation to identify the
entropy production by the driven qubit dynamics. This is the second main result
of the paper.

In section~\ref{sec:laws} we rephrase our results by deriving the form of the
first and second law of thermodynamics. Finally in section~\ref{sec:phonons} we
shortly describe how to extend our model towards a more realistic situation by taking into
account interaction, mediated by phonons, of the calorimeter with the
environment. 


\section{ Modeling}
\label{sec:model} 

The system under consideration consists of a qubit with an energy gap $\hbar \omega_0$ which is 
driven by an external source and which is coupled to an electron system with finite heat capacity 
$C$ (see fig.~\ref{fig:Fig}). The total system is considered to be governed by the Schr\"odinger 
equation with a time dependent  Hamiltonian composed of three parts
\begin{equation*}
	H(t) = H_2(t) + H_e + H_I.
\end{equation*}
Here $H_e$ and $H_I$ denote the free Hamiltonian of the electrons and the interaction 
Hamiltonian between qubit and electrons. The Hamiltonians
\begin{equation*}
	H_2(t) = H_2 + H_d(t) 
\end{equation*} 
acting on the qubit  consists of the free Hamiltonian
$H_2 =  \hbar \omega_0  a^*a$ and  a driving  Hamiltonian of  the form
$H_d(t)  =  \lambda_t(\imath a  -  \imath  a^*)$ varying  smoothly  in
time. Here  we write $a^*$ and  $a$ for the creation  and annihilation
operators of the qubit.


During the process we continuously measure the temperature of the electron
system, henceforth referring to the electrons as calorimeter. We assume that
the number of electrons in the calorimeter is finite but very large and that
the temperature is sufficiently small. Then the Sommerfeld expansion
(see e.g. appendix~C of \cite{AsMe76}) may be used to evaluate the average energy of the
calorimeter as a function of its temperature. Restricting to the leading order
contribution we obtain the relation
\begin{equation}
	\label{eq:Sommerfeld}
	\Delta(T^2) = \frac{4 \, \mathcal E_F}{\pi^2 \, k^2 \, \mathcal N} \; \Delta E
\end{equation}  
between the change of the squared temperature $\Delta(T^2)$ and the change of
the average energy $\Delta E$, where $\mathcal E_F$ is the Fermi energy, $k$ is
the Boltzmann constant, and $\mathcal N$ is the average number of electrons in
the calorimeter.

We assume that the drive changes on much slower time scales than the
interaction between the qubit and the calorimeter. Then the usual Born-Markov
approximation (see \cite{BreuerPetruccione,WiMi09} and \cf
\cite{Bre07,Pel14}) yields an effective master equation for the joint system
consisting of the qubit and the temperature of the calorimeter. The details of
the derivation are outlined in appendix~\ref{app:scales}. For our purposes the
master equation is conveniently written as the It\^o stochastic differential
equation extending the well-known stochastic Schr\"odinger equation (see
\cite{BreuerPetruccione,WiMi09}):
\begin{equation}	\label{eq:stoch_Schr}
	\begin{split}
		\dd\varphi_t
		&= 
		\Bigl( \frac{a_t\varphi_t}{\norm{a_t\varphi_t}} - \varphi_t \Bigr) \, \dd N^+_t 
		+ \Bigl( \frac{a_t^* \varphi_t}{\norm{a_t^*\varphi_t}} - \varphi_t \Bigr) \, \dd N^-_t 
		\\
		&\quad 
		- \tfrac{\imath}{\hbar} \, G_t(T_t) \, \varphi_t \, \dd t  
		\\
		&\quad
		+\tfrac12 \bigl( 
			\gamma^+_t(T_t) \norm{a_t \varphi_t}^2 
			+ \gamma^-_t(T_t) \norm{a_t^*\varphi_t}^2 
		\bigr) \varphi_t \, \dd t
		\\
		\dd(T_t^2) 
		&= \tfrac{4 \mathcal E_F}{\pi^2 k^2 \mathcal N} \, \hbar\omega_t \cdot  (\dd N^+_t - \dd N^-_t) 
	\end{split}
\end{equation}
with the temperature and time dependent non-Hermitian operator 
\begin{align}	\label{eq:non-Herm_gener}
	G_t(T) 
	= 
	H_2(t)
	- \imath \tfrac{\omega_t}{2} \bigl( \gamma^+_t(T) a_t^{} a_t^* + \gamma^-_t(T) a_t^* a_t^{} \bigr).
\end{align} 
In this equation, $\varphi_t$ and $T_t$ are the pure state
of   the   qubit   and   the   measured   temperature   at   time~$t$,
respectively. The operators $a_t$, $a_t^*$ denote the annihilation and
creation operators in the eigenbasis of the qubit Hamiltonian $H_2(t)
= H_2  + H_d(t)$ 
and $\hbar  \omega_t$ is the energy  gap of $H_2(t)$.
The processes $N_t^+$ and $N_t^-$ are Poisson process with independent
increments and rates depending on $\varphi_t$ and $T_t$ via
\begin{align*}
	\expval{\dd N_t^+} &= \gamma^+_t(T_t) \, \norm{a_t \varphi_t}^2 \, \dd t
	\\
	\expval{\dd N_t^-} &= \gamma^-_t(T_t) \, \norm{a_t^* \varphi_t}^2 \, \dd t . 
\end{align*}
The rates $\gamma^\pm_t(T)$ are of the form
\begin{align*}
	\gamma^+_t(T) &= \frac{\gamma}{1 - e^{-\beta(T) \, \omega_t}},
	&
	\gamma^-_t(T) &= \frac{\gamma}{e^{\beta(T) \, \omega_t}-1}
\end{align*} 
for some constant $\gamma$ and the inverse temperature $\beta(T) = 1/kT$. The
particular form of the rates is not essential for all further derivations, but
rather the fact that the rates satisfy the detailed balance condition
\begin{equation*}
	\gamma^+_t(T) = e^{\beta(T) \omega_t} \, \gamma^-_t(T).
\end{equation*}

Before we step into the discussion of the general system, it is expedient to
take a look at the undriven system, \ie, we assume for a moment $H_2(t) = H_2$
for all times.  Let us write $(\varphi, T)$ for the state of the combined
system where the qubit is in the state $\varphi$ up to a phase and the
calorimeter has temperature~$T$. Then, except for an initial period until the
first jump, the  combined system only attains two states jumping between them
back and forth, namely the states $(\up, T)$ for some temperature $T$ and the
state $(\down, T+\Delta T)$ with the next higher temperature level $T + \Delta
T$. This classical two-level system relaxes to a Gibbs equilibrium
distribution. The corresponding Gibbs entropy of the state $(\varphi, T)$ is
given by $\beta(T) \expval{H_2}{\varphi} + S_\equi(T)$, where $S_\equi$ denotes
the \emph{equilibrium entropy} of the calorimeter:
\begin{equation*}
	S_\equi(T ) := - \ln \gamma_t^-(T).
\end{equation*}

\section{Measurements and Their Probabilities}
\label{sec:path_distr}

For  the  system  described  above we  study  a  concrete  measurement
protocol specified  as follows: We  fix a time  interval $[t_\initial,
t_\final]$ and we assume that the  drive vanishes at the beginning and
the  end   of  the   interval,  \ie,   $\lambda_{t_\initial}  =   0  =
\lambda_{t_\final}$. During the interval we proceed as follows:
\begin{enumerate}
\item At initial  time $t_\initial$, we prepare the  qubit randomly in
one   of    its   eigenstates   $\varphi_\initial   =    \ket\up$   or
$\varphi_\initial = \ket\down$. Equivalently, we assume that the qubit
is directly measured in this basis.
\item  During   the  time   interval  we  continuously   measure  the
temperature  of   the  calorimeter,   which  provides   a  temperature
trajectory $T_t$.
\item At  final time $t_f$,  we directly  measure the energy  of qubit
resulting    in    a    state   $\varphi_\final    =    \ket\up$    or
\mbox{$\varphi_\final = \ket\down$}.
\end{enumerate}

In order to make the distribution of the temperature trajectory explicit, let us fix a trajectory until time $t_\initial \le t \le t_\final$. We write $t_\initial = t_0 < t_1 < \dots < t_n < t_{n+1} = t$ for the partition of the time interval such that the temperature jumps at times $t_k$ for $1 \le k \le n$ and remains constant at temperature $T_k$ during the interval $(t_{k-1}, t_k)$ for $1\le k \le n+1$. Then the state of the qubit at time $t$ is given by renormalizing the vector
\begin{align} 	
	\notag
	\tilde\varphi_t
	&= 
	U_{t,t_n}(T_n) \, \sqrt{\gamma_n} a_n \dots  U_{t_2,t_1}(T_1) \, \sqrt{\gamma_1} a_1 \, U_{t_1,t_0}(T_0) \varphi_\initial
	\\
	&=
	\Bigl( \prod_{k=1}^n \gamma_k \Bigr)^{1/2} U_{t,t_n}(T_n) \dots a_1 U_{t_1,t_0}(T_0) \varphi_\initial,
	\label{eq:evol_vector}
\end{align}  
where for sake of brevity we put $\gamma_k := \gamma_{t_k}^+(T_{k-1})$, $a_k :=
a_t$ if $T_k > T_{k-1}$ and otherwise we put $\gamma_k :=
\gamma_{t_k}^-(T_{k-1})$, $a_k := a_t^*$. The operators $U_{t,s}(T)$ denote the
two-parameter semigroup acting on the Hilbert space of the qubit given by the
solution of $U_{s,s}(T) = 1$ and $\dv{t} U_{t,s}(T) = -\tfrac{\imath}{\hbar}
G_t(T) \, U_{t,s}(T)$ for all $s \le t$. For a trajectory of the full time
horizon $[t_\initial,t_\final]$ the conditional distribution of a temperature
trajectory $(T_t)_t$ and a final measurement result $\varphi_\final$ is hence
given by
\begin{multline}
	\label{eq:prob}
	\mathbb P(T_t, \varphi_\final \; | \; \varphi_\initial)
	:= \abs{\braket{\varphi_\final\;}{\;\tilde\varphi_{t_\final}}}^2
	\\
	= \Bigl( \prod_{k=1}^n \gamma_k \Bigr) \,  
	\abs{ 
		\bra{\varphi_\final} U_{t_f,t_n}(T_n) \dots U_{t_1, t_0}(T_0) \ket{\varphi_\initial}
	}^2.
\end{multline}

\section{Fluctuation Relation and Entropy Production}
\label{sec:fr}

In order to derive a fluctuation relation for the investigated process we
choose a natural time reversal. For distinction we henceforth refer to the
process described in section~\ref{sec:model}, equation~\eqref{eq:stoch_Schr},
as the forward process. The dynamics of the reversed process is also determined
by equation~\eqref{eq:stoch_Schr} where we replace $G_t(T)$ by the operator
$G_t^R(T)$ given by
\begin{equation*}
	G^R_{t_\final - t}(T) 
	\! =  \!
	-H_2(t)
	-\imath \tfrac{\omega_t}{2} \bigl( 
		\gamma^+_t(T) a_t^{} a_t^* + \gamma^-_t(T) a_t^* a_t^{} 
	\bigr).
\end{equation*} 
The resulting equation may equivalently be derived by inverting the initial
Schr\"odinger equation of the combined quantum system consisting of qubit and
calorimeter and performing the same approximation described in
section~\ref{sec:model}. We apply the same protocol to the time reversed
process, that is, at time $t_\initial$ we prepare the qubit, we observe the
temperature during the time interval $[t_\initial, t_f]$, and we measure the
qubit directly at final time $t_f$. Proceeding as in
section~\ref{sec:path_distr} we may analogously derive the distribution on the
path space for the reversed process.

We now compare  a trajectory of the forward process,  given by initial
and  final qubit  state $\varphi_\initial$,  $\varphi_\final$ and  the
temperature $T_t$,  with the trajectory  of the reversed  process that
runs backward in time, \ie, the trajectory that starts with a qubit in
state $\varphi_\final$, yields  the temperatures $T_{t_\final-t}$, and
results in $\varphi_\initial$ in the final measurement.

In order to compare the path distribution of the forward process $\prob$ and
the reversed process $\prob^R$, we may observe the following two facts. First,
for each jump in a trajectory of the forward process there is a jump in the
opposite direction in the corresponding trajectory of the reversed process.
However, the jumping rates are different. Second, the evolution in between
temperature jumps is governed by the two-parameter semigroups $U_{t,s}(T)$ with
generator $-\tfrac{\imath}{\hbar} G_t(T)$ for the forward process and by
$U_{t,s}^R(T)$ with generator $-\tfrac{\imath}{\hbar} G^R_t(T)$ for the
reversed process, respectively. Since $G^R_{t_f-t}(T) = -G_t(T)^*$, the
associated semigroups satisfy
\begin{equation*}
	U_{t,s}^R(T) = U_{t_f-s, t_f-t}(T)^*
\end{equation*}   
for all $s \le t$. We may now compare the expression \eqref{eq:evol_vector} and
\eqref{eq:prob} for the density of the forward process with the corresponding
expressions for density of the backward process. Gathering the coefficients
then immediately yields
\begin{equation}
	\label{eq:Radon-Nikodym}
	\prob(T_t, \varphi_\final \;|\; \varphi_\initial) 
	= 
	e^J \; \prob^R(T_{t_\final -t}, \varphi_\initial \;|\; \varphi_\final).
\end{equation}
Here $J$ is the function that, on a path of the forward process, is given by
\begin{equation*}
	J := \sum_{k=1}^n \bigl( \ln \gamma^{x_k}_{t_k}(T_{k-1}) - \ln \gamma^{-x_k}_{t_k}(T_k) \bigr),
\end{equation*} 
where  as before $T_\initial  = T_0, T_1,  \dots, T_n=
T_\final$ is the observed temperature  trajectory, $t_\initial < t_1 <
\dots, t_n  < t_f$  
are the  jumping times, and  $x_1, \dots,  x_n \in
\{+1,-1\}$  are the  jumping directions  $x_k :=  \dd N_{t_k}^+  - \dd
N_{t_k}^-$. 
Due to  the detailed balanced condition for  the rates, we
may write $J$ as
\begin{align*}
	J 
	&= \sum_{k=1}^n \bigl( \Delta S_{\equi,k} - \beta_k Q_k \bigr),
\end{align*} 
where, for each jump $k$, we write $\Delta S_{\equi,k} := S_\equi(t_k, T_k) -S_\equi(t_k, T_{k-1})$ for the change of free entropy in the calorimeter, $Q_k := \hbar \omega_{t_k} x_k$ is the heat emitted by the qubit into the calorimeter, and where we put $\beta_k := \beta(T_{k-1})$ if $x_k = 1$ and $\beta_k := \beta(T_k)$ otherwise. (The case distinction is due to the It\^o convention in equation~\eqref{eq:stoch_Schr}.)

We may interpret $J$ as an entropy flux from the qubit to the calorimeter.
Another contribution to the entropy production is the entropy of the initial
preparation and the final measurement of the qubit given by $S_{\initial /
\final}(\varphi) = -\ln \prob_{\initial / \final}(\varphi)$, where
$\prob_{\initial / \final}(\varphi)$ is the probability that the state is
initially prepared in the state $\varphi \in \{\up, \down\}$ or, respectively,
finally observed in $\varphi$. The entropy production of the qubit along a
trajectory is then given~by
\begin{equation*}
	\sigma := S_\final - S_\initial + J.
\end{equation*} 
Equation~\eqref{eq:Radon-Nikodym} immediately implies $\expval{e^{-\sigma}} =
1$ for the expectation value with 
respect to the forward process. In particular, it follows $\expval{\sigma} \ge 0$.

The derived entropy production can be compared with the approaches to entropy
production by Horowitz and Parrondo~\cite{HoPa13} and by Breuer~\cite{Bre03} in
the limit of an infinite calorimeter, \ie, when the heat capacity of the
calorimeter becomes infinite. In this case the resulting temperature change is negligible during the
driving period, and the work and heat measurement strategy needs to be reconsidered. 
Moreover, in \cite{Bre03} the drive is considered as a small perturbation of
the free evolution in the sense that the time dependence of the energy gap of
the qubit Hamiltonian is neglected. In this case the entropy flux takes the
form
\begin{equation*}
	J 
	= -\sum_{k=1}^n \beta \hbar \omega_0 x_k
	= -\beta \hbar \omega_0 (N^+_{t_\final} - N^-_{t_\final}),
\end{equation*} 
where $\beta = 1/kT$ denotes the inverse temperature and where we assume
without loss of generality that the Poisson processes start at $N^+_0 = N^-_0 =
0$. 
Taking the expectation value we recover the entropy flux
of~\cite{Bre03,HoPa13}. Due to the final measurement of the qubit, the
expectation of the entropy production $\expval{\sigma}$ is in general larger or
equal to the entropy production in~\cite{Bre03}.

\section{First and Second Law of Thermodynamics}
\label{sec:laws}

For the first law we concentrate on the qubit. Suppose at time $t$ the
state of  the qubit is  given by  $\varphi_t$ and the  calorimeter has
temperature~$T$. Then the expectation value of the energy of the qubit
is given by
\begin{equation*}
	\mathcal E(t) := \expval{H_2(t)}{\varphi_t}
\end{equation*}   
with the qubit Hamiltonian $H_2(t) = H_2 + H_d(t)$. Following
\cite{PuWo78,Ali79}, the work per unit of time performed on the qubit at time
$t$ is given~by
\begin{equation*}
	\mathcal W(t) := \expval**{\dv{t} H_2(t)}{\varphi_t} = \expval**{\dv{t} H_d(t)}{\varphi_t}.
\end{equation*}
Furthermore, the heat per unit of time emitted by the qubit into the
calorimeter at time $t$ is given~by 
\begin{equation*}
	\dd \mathcal Q(t) := \hbar \omega_t (\dd N_t^+ - \dd N_t^-).
\end{equation*}
Indeed a straightforward computation with \eqref{eq:stoch_Schr} yields the 
stochastic form of the first law of thermodynamics in expectation value:
\begin{equation*}
	\expval{\dd \mathcal E(t)}_{\varphi_t}
	= \mathcal W_t \; \dd t - \expval{\dd \mathcal Q(t)}_{\varphi_t},
\end{equation*}
where $\expval{\, \cdot \,}_{\varphi_t}$ denotes the expectation value
conditioned on the qubit being in state $\varphi_t$.

The second law of thermodynamics may be derived in the limit of an infinite
heat capacity. In this case the calorimeter forms an infinite heat bath for the
qubit with constant temperature~$T$. The entropy flux $J$ then simplifies
to $J = -\beta Q$, where for each trajectory $Q := \sum_{k=1}^n Q_k$ is the
total heat emitted by the qubit. The total entropy change in the qubit along a
trajectory is hence given by
\begin{equation*}
	\Delta S := S_\final -S_\initial = \sigma + \beta Q. 
\end{equation*}
Since $\expval{\sigma} \ge 0$, taking the expectation value yields the familiar
form of the second law of thermodynamics,
\begin{equation*}
	\expval{\Delta S} \ge +\beta \expval{Q}.
\end{equation*}

\section{Including  a Phonon  Bath} 
\label{sec:phonons}

The model studied so far can be easily extended to the case where the
calorimeter is emerged into a phonon heat bath. We assume that the interaction
between calorimeter and the phonon bath is weaker but much faster than the
interaction between the qubit and calorimeter. Then the energy exchange between
calorimeter and phonons only depends on the respective temperatures. For a
calorimeter made of ordinary metal, for instance, we have
\begin{equation}
	\label{eq:phonon_heat}
	\dot Q_{ep} = \Sigma_{ep} \, ( T^5 - T_p^5),
\end{equation}  
where $T$ is the temperature of the electron calorimeter, $T_p$ is the
temperature of the phonon bath, $Q_{ep}$ is the heat emitted by the calorimeter
into the phonon bath, and $\Sigma_{ep}$ is a material dependent constant (see
\cite{WeUrCl94}). For materials with a more sophisticated structure, \eg, thin
metallic layers,  the law \eqref{eq:phonon_heat} may be different (see
\cite{CoAn16}). For simplicity we assume that the  energy exchange is
deterministic. Using as before the approximation provided by the Sommerfeld
expansion to relate energy changes and temperature changes in the calorimeter
via eq.~\eqref{eq:Sommerfeld}, we find that the second equation of
\eqref{eq:stoch_Schr} has to be replaced~by
\begin{equation*}
	\dd (T_t^2)
	= \tfrac{4 \mathcal E_F}{\pi^2 k^2 \mathcal N} \Bigl( 
		\hbar\omega_t  (\dd N^+_t - \dd N^-_t) +  \Sigma_{ep} (T_p^5 - T_t^5) \; \dd t 
	\Bigr).
\end{equation*}

\section{Conclusions}

Summarizing we have established a mathematical model for the dynamics of an
externally driven qubit interacting with a continuously monitored, finite size
calorimeter. The resulting quantum stochastic evolution \eqref{eq:stoch_Schr}
extends the usual stochastic Schr\"odinger equation by incorporating the
changing temperature of the calorimeter and its back-action on the measurement
process. We derived an explicit expression for the entropy production and
demonstrated a corresponding fluctuation relation. As a first application we verified
the first law of thermodynamics and proved the second law of thermodynamics in
reasonable limiting cases. Finally, we indicated how the model can be extended
easily to comprise a common phonon bath.  

The model described here is intended as a first approach to the dynamics
offering a simple presentation. Experiments suggest that a more realistic model
should treat the phonon bath stochastically as well. Such a model is currently
investigated numerically \cite{Don16} in order to analyze the fluctuating
thermodynamic quantities beyond the fluctuation relation. In some cases an
analytic treatment seems feasible, too. Currently designed experiments address
the fluctuation relation derived here. In this way the implications of the
finite size of the calorimeter are tested (\cf \cite{SuSaPeAnAl14}) and the
non-Markovianity of the dynamics can be quantified.

\section{Acknowledgements}
The authors would like to thank Rapha\"el Ch\'etrite, Yuri Galperin, Samu
Suomela, and Brecht Donvil for fruitful discussions and comments. This work has
been supported by the Academy of Finland (project no.~272218, 284594, 271983).

\appendix

\section{Separating Time Scales}
\label{app:scales} 

The fastest time scale in the interaction between qubit and calorimeter is the electron-electron interaction in the calorimeter. This interaction is much faster than any other time-scale involved and does not effect the average energy. Therefore, we treat the self-interaction as an average effect in the van-Hove weak coupling limit  \cite{vHo54,Dav74,Dav76}. On the relevant time scales the calorimeter then can be assumed to be always in a well-defined temperature state characterized by its mean energy.

The second fastest time scale is the interaction between the qubit and the
calorimeter. The drive changes on much slower time scale. Then the
qubit-calorimeter interaction can be treated by the usual Born-Markov
approximation (see \cite[Sec.~3.3, Sec.~8.4]{BreuerPetruccione} or
\cite[Sec.~3.2]{WiMi09}), where the drive is considered to be constant during
the interaction. Since we only study  energy changes in the calorimeter, we may
for this purpose regard the calorimeter as free fermions with corresponding
free Hamiltonian $H_e = \sum_k \epsilon_k \, c_k^* c_k^{}$ and we may take the
interaction Hamiltonian of the form
\begin{equation*}
	H_I = \sum_{k \neq \ell} g_{k,\ell} (a + a^*) c_k^* c_\ell,
\end{equation*}
where the sums are taken over all (pairs of) fermionic modes with corresponding
energy $\epsilon_k$ and with some interaction coefficients $g_{k,\ell} =
\overline{g_{\ell, k}}$. 

It is expedient to write the drive in the form 
\begin{equation*}
	H_d(t) = \tfrac12 \hbar \omega_0 \tan \theta_t \, (\imath a - \imath a^*) 
\end{equation*}
for some smooth parameter $-\pi/2 < \theta_t < \pi/2$. For this choice of
parameters the total Hamiltonian of the qubit $H_2(t) = H_2 + H_d(t)$ is
diagonalized by the unitary operator
\begin{equation*}
	U_t := \begin{pmatrix}
		\cos(\theta_t/2) & -\imath \sin(\theta_t/2)
		\\
		-\imath \sin(\theta_t/2) & \cos(\theta_t/2)
	\end{pmatrix}.
\end{equation*}  
with generator $H_A(t) := \imath \dot U_t U_t^*  = -\tfrac12 \dot
\theta_t(a+a^*)$. It is worth noting that the field operator $a+a^*$ commutes
with $H_A(t)$ and hence satisfies $U_t(a+a^*)U_t^* = a+a^*$. Passing to the
corresponding interaction picture hence does not effect the interaction
Hamiltonian.  For the derivation of the master equation we treat the drive
adiabatically, that is, we neglect the contribution of $H_A(t)$. For the form
of the master equation including the computation of the rates we refer to \eg
\cite[Sec.~6]{BreuerPetruccione} (see also \cite{SuSaPeAnAl14,PeSuGa16}),
where similar models are treated.

\addcontentsline{toc}{section}{Bibliography}
\bibliography{qmeasurement}

\begin{thebibliography}{56}%
\makeatletter
\providecommand \@ifxundefined [1]{%
 \@ifx{#1\undefined}
}%
\providecommand \@ifnum [1]{%
 \ifnum #1\expandafter \@firstoftwo
 \else \expandafter \@secondoftwo
 \fi
}%
\providecommand \@ifx [1]{%
 \ifx #1\expandafter \@firstoftwo
 \else \expandafter \@secondoftwo
 \fi
}%
\providecommand \natexlab [1]{#1}%
\providecommand \enquote  [1]{``#1''}%
\providecommand \bibnamefont  [1]{#1}%
\providecommand \bibfnamefont [1]{#1}%
\providecommand \citenamefont [1]{#1}%
\providecommand \href@noop [0]{\@secondoftwo}%
\providecommand \href [0]{\begingroup \@sanitize@url \@href}%
\providecommand \@href[1]{\@@startlink{#1}\@@href}%
\providecommand \@@href[1]{\endgroup#1\@@endlink}%
\providecommand \@sanitize@url [0]{\catcode `\\12\catcode `\$12\catcode
  `\&12\catcode `\#12\catcode `\^12\catcode `\_12\catcode `\%12\relax}%
\providecommand \@@startlink[1]{}%
\providecommand \@@endlink[0]{}%
\providecommand \url  [0]{\begingroup\@sanitize@url \@url }%
\providecommand \@url [1]{\endgroup\@href {#1}{\urlprefix }}%
\providecommand \urlprefix  [0]{URL }%
\providecommand \Eprint [0]{\href }%
\providecommand \doibase [0]{http://dx.doi.org/}%
\providecommand \selectlanguage [0]{\@gobble}%
\providecommand \bibinfo  [0]{\@secondoftwo}%
\providecommand \bibfield  [0]{\@secondoftwo}%
\providecommand \translation [1]{[#1]}%
\providecommand \BibitemOpen [0]{}%
\providecommand \bibitemStop [0]{}%
\providecommand \bibitemNoStop [0]{.\EOS\space}%
\providecommand \EOS [0]{\spacefactor3000\relax}%
\providecommand \BibitemShut  [1]{\csname bibitem#1\endcsname}%
\let\auto@bib@innerbib\@empty
\bibitem [{\citenamefont {Landauer}(1961)}]{Lan61}%
  \BibitemOpen
  \bibfield  {author} {\bibinfo {author} {\bibfnamefont {R.}~\bibnamefont
  {Landauer}},\ }\href
  {http://researchweb.watson.ibm.com/journal/rd/053/ibmrd0503C.pdf} {\bibfield
  {journal} {\bibinfo  {journal} {IBM Journal of Research and Development}\
  }\textbf {\bibinfo {volume} {5}},\ \bibinfo {pages} {183} (\bibinfo {year}
  {1961})}\BibitemShut {NoStop}%
\bibitem [{\citenamefont {Bekenstein}(1981)}]{Bek81}%
  \BibitemOpen
  \bibfield  {author} {\bibinfo {author} {\bibfnamefont {J.~D.}\ \bibnamefont
  {Bekenstein}},\ }\href {\doibase 10.1103/physrevlett.46.623} {\bibfield
  {journal} {\bibinfo  {journal} {Physical Review Letters}\ }\textbf {\bibinfo
  {volume} {46}},\ \bibinfo {pages} {623} (\bibinfo {year} {1981})}\BibitemShut
  {NoStop}%
\bibitem [{\citenamefont {Lloyd}(2000)}]{LLo00}%
  \BibitemOpen
  \bibfield  {author} {\bibinfo {author} {\bibfnamefont {S.}~\bibnamefont
  {Lloyd}},\ }\href {\doibase 10.1038/35023282} {\bibfield  {journal} {\bibinfo
   {journal} {Nature}\ }\textbf {\bibinfo {volume} {406}},\ \bibinfo {pages}
  {1047} (\bibinfo {year} {2000})}\BibitemShut {NoStop}%
\bibitem [{\citenamefont {Leff}\ and\ \citenamefont {Rex}(2003)}]{LeffRex03}%
  \BibitemOpen
  \bibfield  {author} {\bibinfo {author} {\bibfnamefont {H.~S.}\ \bibnamefont
  {Leff}}\ and\ \bibinfo {author} {\bibfnamefont {A.~F.}\ \bibnamefont {Rex}},\
  }\href@noop {} {\emph {\bibinfo {title} {{Maxwell's demon 2: entropy,
  classical and quantum information, computing}}}},\ \bibinfo {edition} {2nd}\
  ed.\ (\bibinfo  {publisher} {Institute of Physics Publishing},\ \bibinfo
  {year} {2003})\ p.\ \bibinfo {pages} {485}\BibitemShut {NoStop}%
\bibitem [{\citenamefont {Schmidt}\ \emph {et~al.}(2003)\citenamefont
  {Schmidt}, \citenamefont {Yung},\ and\ \citenamefont {Cleland}}]{ScYuCl03}%
  \BibitemOpen
  \bibfield  {author} {\bibinfo {author} {\bibfnamefont {D.~R.}\ \bibnamefont
  {Schmidt}}, \bibinfo {author} {\bibfnamefont {C.~S.}\ \bibnamefont {Yung}}, \
  and\ \bibinfo {author} {\bibfnamefont {A.~N.}\ \bibnamefont {Cleland}},\
  }\href {\doibase 10.1063/1.1597983} {\bibfield  {journal} {\bibinfo
  {journal} {Applied Physics Letters}\ }\textbf {\bibinfo {volume} {83}},\
  \bibinfo {pages} {1002} (\bibinfo {year} {2003})}\BibitemShut {NoStop}%
\bibitem [{\citenamefont {Schmidt}\ \emph {et~al.}(2004)\citenamefont
  {Schmidt}, \citenamefont {Schoelkopf},\ and\ \citenamefont
  {Cleland}}]{ScScCl04}%
  \BibitemOpen
  \bibfield  {author} {\bibinfo {author} {\bibfnamefont {D.~R.}\ \bibnamefont
  {Schmidt}}, \bibinfo {author} {\bibfnamefont {R.~J.}\ \bibnamefont
  {Schoelkopf}}, \ and\ \bibinfo {author} {\bibfnamefont {A.~N.}\ \bibnamefont
  {Cleland}},\ }\href {\doibase 10.1103/physrevlett.93.045901} {\bibfield
  {journal} {\bibinfo  {journal} {Physical Review Letters}\ }\textbf {\bibinfo
  {volume} {93}},\ \bibinfo {pages} {045901} (\bibinfo {year}
  {2004})}\BibitemShut {NoStop}%
\bibitem [{\citenamefont {Gasparinetti}\ \emph {et~al.}(2015)\citenamefont
  {Gasparinetti}, \citenamefont {Viisanen}, \citenamefont {Saira},
  \citenamefont {Faivre}, \citenamefont {Arzeo}, \citenamefont {Meschke},\ and\
  \citenamefont {Pekola}}]{GaViSaFaArMePe15}%
  \BibitemOpen
  \bibfield  {author} {\bibinfo {author} {\bibfnamefont {S.}~\bibnamefont
  {Gasparinetti}}, \bibinfo {author} {\bibfnamefont {K.~L.}\ \bibnamefont
  {Viisanen}}, \bibinfo {author} {\bibfnamefont {O.-P.}\ \bibnamefont {Saira}},
  \bibinfo {author} {\bibfnamefont {T.}~\bibnamefont {Faivre}}, \bibinfo
  {author} {\bibfnamefont {M.}~\bibnamefont {Arzeo}}, \bibinfo {author}
  {\bibfnamefont {M.}~\bibnamefont {Meschke}}, \ and\ \bibinfo {author}
  {\bibfnamefont {J.~P.}\ \bibnamefont {Pekola}},\ }\href {\doibase
  10.1103/physrevapplied.3.014007} {\bibfield  {journal} {\bibinfo  {journal}
  {Physical Review Applied}\ }\textbf {\bibinfo {volume} {3}},\ \bibinfo
  {pages} {014007} (\bibinfo {year} {2015})},\ \Eprint
  {http://arxiv.org/abs/1405.7568} {arXiv:1405.7568 [cond-mat.mes-hall]}
  \BibitemShut {NoStop}%
\bibitem [{\citenamefont {Viisanen}\ \emph {et~al.}(2015)\citenamefont
  {Viisanen}, \citenamefont {Suomela}, \citenamefont {Gasparinetti},
  \citenamefont {Saira}, \citenamefont {Ankerhold},\ and\ \citenamefont
  {Pekola}}]{ViSuGaSaAnPe15}%
  \BibitemOpen
  \bibfield  {author} {\bibinfo {author} {\bibfnamefont {K.~L.}\ \bibnamefont
  {Viisanen}}, \bibinfo {author} {\bibfnamefont {S.}~\bibnamefont {Suomela}},
  \bibinfo {author} {\bibfnamefont {S.}~\bibnamefont {Gasparinetti}}, \bibinfo
  {author} {\bibfnamefont {O.-P.}\ \bibnamefont {Saira}}, \bibinfo {author}
  {\bibfnamefont {J.}~\bibnamefont {Ankerhold}}, \ and\ \bibinfo {author}
  {\bibfnamefont {J.~P.}\ \bibnamefont {Pekola}},\ }\href {\doibase
  10.1088/1367-2630/17/5/055014} {\bibfield  {journal} {\bibinfo  {journal}
  {New Journal of Physics}\ }\textbf {\bibinfo {volume} {17}},\ \bibinfo
  {pages} {055014} (\bibinfo {year} {2015})},\ \Eprint
  {http://arxiv.org/abs/1412.7322} {arXiv:1412.7322 [cond-mat.stat-mech]}
  \BibitemShut {NoStop}%
\bibitem [{\citenamefont {Giazotto}\ \emph {et~al.}(2006)\citenamefont
  {Giazotto}, \citenamefont {Heikkil\"a}, \citenamefont {Luukanen},
  \citenamefont {Savin},\ and\ \citenamefont {Pekola}}]{GiHeLuSaPe06}%
  \BibitemOpen
  \bibfield  {author} {\bibinfo {author} {\bibfnamefont {F.}~\bibnamefont
  {Giazotto}}, \bibinfo {author} {\bibfnamefont {T.~T.}\ \bibnamefont
  {Heikkil\"a}}, \bibinfo {author} {\bibfnamefont {A.}~\bibnamefont
  {Luukanen}}, \bibinfo {author} {\bibfnamefont {A.~M.}\ \bibnamefont {Savin}},
  \ and\ \bibinfo {author} {\bibfnamefont {J.~P.}\ \bibnamefont {Pekola}},\
  }\href {\doibase 10.1103/revmodphys.78.217} {\bibfield  {journal} {\bibinfo
  {journal} {Review of Modern Physics}\ }\textbf {\bibinfo {volume} {78}},\
  \bibinfo {pages} {217} (\bibinfo {year} {2006})}\BibitemShut {NoStop}%
\bibitem [{\citenamefont {Pekola}\ \emph {et~al.}(2013)\citenamefont {Pekola},
  \citenamefont {Solinas}, \citenamefont {Shnirman},\ and\ \citenamefont
  {Averin}}]{PeSoShAv13}%
  \BibitemOpen
  \bibfield  {author} {\bibinfo {author} {\bibfnamefont {J.~P.}\ \bibnamefont
  {Pekola}}, \bibinfo {author} {\bibfnamefont {P.}~\bibnamefont {Solinas}},
  \bibinfo {author} {\bibfnamefont {A.}~\bibnamefont {Shnirman}}, \ and\
  \bibinfo {author} {\bibfnamefont {D.~V.}\ \bibnamefont {Averin}},\ }\href
  {\doibase 10.1088/1367-2630/15/11/115006} {\bibfield  {journal} {\bibinfo
  {journal} {New Journal of Physics}\ }\textbf {\bibinfo {volume} {15}},\
  \bibinfo {pages} {115006} (\bibinfo {year} {2013})},\ \Eprint
  {http://arxiv.org/abs/1212.5808} {arXiv:1212.5808 [cond-mat.stat-mech]}
  \BibitemShut {NoStop}%
\bibitem [{\citenamefont {Clarke}\ and\ \citenamefont
  {Wilhelm}(2008)}]{ClWi08}%
  \BibitemOpen
  \bibfield  {author} {\bibinfo {author} {\bibfnamefont {J.}~\bibnamefont
  {Clarke}}\ and\ \bibinfo {author} {\bibfnamefont {F.~K.}\ \bibnamefont
  {Wilhelm}},\ }\href {\doibase 10.1038/nature07128} {\bibfield  {journal}
  {\bibinfo  {journal} {Nature}\ }\textbf {\bibinfo {volume} {453}},\ \bibinfo
  {pages} {1031–1042} (\bibinfo {year} {2008})}\BibitemShut {NoStop}%
\bibitem [{\citenamefont {Averin}\ \emph {et~al.}(1984)\citenamefont {Averin},
  \citenamefont {Zorin},\ and\ \citenamefont {Likharev}}]{AvZoLi84}%
  \BibitemOpen
  \bibfield  {author} {\bibinfo {author} {\bibfnamefont {D.~V.}\ \bibnamefont
  {Averin}}, \bibinfo {author} {\bibfnamefont {A.~B.}\ \bibnamefont {Zorin}}, \
  and\ \bibinfo {author} {\bibfnamefont {K.~K.}\ \bibnamefont {Likharev}},\
  }\href {http://www.jetp.ac.ru/cgi-bin/e/index/e/61/2/p407?a=list} {\bibfield
  {journal} {\bibinfo  {journal} {Journal of Experimental and Theoretical
  Physics (ZhETF, Vol. 88, No. 2, p. 692)}\ }\textbf {\bibinfo {volume} {61}},\
  \bibinfo {pages} {407} (\bibinfo {year} {1984})}\BibitemShut {NoStop}%
\bibitem [{\citenamefont {B\"uttiker}(1987)}]{But87}%
  \BibitemOpen
  \bibfield  {author} {\bibinfo {author} {\bibfnamefont {M.}~\bibnamefont
  {B\"uttiker}},\ }\href {\doibase 10.1103/physrevb.36.3548} {\bibfield
  {journal} {\bibinfo  {journal} {Physical Review B}\ }\textbf {\bibinfo
  {volume} {36}},\ \bibinfo {pages} {3548} (\bibinfo {year}
  {1987})}\BibitemShut {NoStop}%
\bibitem [{\citenamefont {Makhlin}\ \emph {et~al.}(1999)\citenamefont
  {Makhlin}, \citenamefont {Sch\"on},\ and\ \citenamefont
  {Shnirman}}]{MaScSh99}%
  \BibitemOpen
  \bibfield  {author} {\bibinfo {author} {\bibfnamefont {Y.}~\bibnamefont
  {Makhlin}}, \bibinfo {author} {\bibfnamefont {G.}~\bibnamefont {Sch\"on}}, \
  and\ \bibinfo {author} {\bibfnamefont {A.}~\bibnamefont {Shnirman}},\ }\href
  {\doibase 10.1038/18613} {\bibfield  {journal} {\bibinfo  {journal} {Nature}\
  }\textbf {\bibinfo {volume} {398}},\ \bibinfo {pages} {305–307} (\bibinfo
  {year} {1999})}\BibitemShut {NoStop}%
\bibitem [{\citenamefont {Koch}\ \emph {et~al.}(2007)\citenamefont {Koch},
  \citenamefont {Yu}, \citenamefont {Gambetta}, \citenamefont {Houck},
  \citenamefont {Schuster}, \citenamefont {Majer}, \citenamefont {Blais},
  \citenamefont {Devoret}, \citenamefont {Girvin},\ and\ \citenamefont
  {Schoelkopf}}]{KoYuGaHoSc07}%
  \BibitemOpen
  \bibfield  {author} {\bibinfo {author} {\bibfnamefont {J.}~\bibnamefont
  {Koch}}, \bibinfo {author} {\bibfnamefont {T.~M.}\ \bibnamefont {Yu}},
  \bibinfo {author} {\bibfnamefont {J.}~\bibnamefont {Gambetta}}, \bibinfo
  {author} {\bibfnamefont {A.~A.}\ \bibnamefont {Houck}}, \bibinfo {author}
  {\bibfnamefont {D.~I.}\ \bibnamefont {Schuster}}, \bibinfo {author}
  {\bibfnamefont {J.}~\bibnamefont {Majer}}, \bibinfo {author} {\bibfnamefont
  {A.}~\bibnamefont {Blais}}, \bibinfo {author} {\bibfnamefont {M.~H.}\
  \bibnamefont {Devoret}}, \bibinfo {author} {\bibfnamefont {S.~M.}\
  \bibnamefont {Girvin}}, \ and\ \bibinfo {author} {\bibfnamefont {R.~J.}\
  \bibnamefont {Schoelkopf}},\ }\href {\doibase 10.1103/physreva.76.042319}
  {\bibfield  {journal} {\bibinfo  {journal} {Physical Review A}\ }\textbf
  {\bibinfo {volume} {76}},\ \bibinfo {pages} {042319} (\bibinfo {year}
  {2007})},\ \Eprint {http://arxiv.org/abs/cond-mat/0703002}
  {arXiv:cond-mat/0703002 [cond-mat.mes-hall]} \BibitemShut {NoStop}%
\bibitem [{\citenamefont {Schreier}\ \emph {et~al.}(2008)\citenamefont
  {Schreier}, \citenamefont {Houck}, \citenamefont {Koch}, \citenamefont
  {Schuster}, \citenamefont {Johnson}, \citenamefont {Chow}, \citenamefont
  {Gambetta}, \citenamefont {Majer}, \citenamefont {Frunzio}, \citenamefont
  {Devoret}, \citenamefont {Girvin},\ and\ \citenamefont {J.}}]{ScHoKoScJo08}%
  \BibitemOpen
  \bibfield  {author} {\bibinfo {author} {\bibfnamefont {J.~A.}\ \bibnamefont
  {Schreier}}, \bibinfo {author} {\bibfnamefont {A.~A.}\ \bibnamefont {Houck}},
  \bibinfo {author} {\bibfnamefont {J.}~\bibnamefont {Koch}}, \bibinfo {author}
  {\bibfnamefont {D.~I.}\ \bibnamefont {Schuster}}, \bibinfo {author}
  {\bibfnamefont {B.~R.}\ \bibnamefont {Johnson}}, \bibinfo {author}
  {\bibfnamefont {J.~M.}\ \bibnamefont {Chow}}, \bibinfo {author}
  {\bibfnamefont {J.~M.}\ \bibnamefont {Gambetta}}, \bibinfo {author}
  {\bibfnamefont {J.}~\bibnamefont {Majer}}, \bibinfo {author} {\bibfnamefont
  {L.}~\bibnamefont {Frunzio}}, \bibinfo {author} {\bibfnamefont {M.~H.}\
  \bibnamefont {Devoret}}, \bibinfo {author} {\bibfnamefont {S.~M.}\
  \bibnamefont {Girvin}}, \ and\ \bibinfo {author} {\bibfnamefont {S.~R.}\
  \bibnamefont {J.}},\ }\href {\doibase 10.1103/physrevb.77.180502} {\bibfield
  {journal} {\bibinfo  {journal} {Physical Review B}\ }\textbf {\bibinfo
  {volume} {77}},\ \bibinfo {pages} {180502(R)} (\bibinfo {year} {2008})},\
  \Eprint {http://arxiv.org/abs/0712.3581} {arXiv:0712.3581
  [cond-mat.mes-hall]} \BibitemShut {NoStop}%
\bibitem [{\citenamefont {Bouchiat}\ \emph {et~al.}(1998)\citenamefont
  {Bouchiat}, \citenamefont {Vion}, \citenamefont {Joyez}, \citenamefont
  {Esteve},\ and\ \citenamefont {Devoret}}]{BoViJoEsDe98}%
  \BibitemOpen
  \bibfield  {author} {\bibinfo {author} {\bibfnamefont {V.}~\bibnamefont
  {Bouchiat}}, \bibinfo {author} {\bibfnamefont {D.}~\bibnamefont {Vion}},
  \bibinfo {author} {\bibfnamefont {P.}~\bibnamefont {Joyez}}, \bibinfo
  {author} {\bibfnamefont {D.}~\bibnamefont {Esteve}}, \ and\ \bibinfo {author}
  {\bibfnamefont {M.~H.}\ \bibnamefont {Devoret}},\ }\href {\doibase
  10.1238/physica.topical.076a00165} {\bibfield  {journal} {\bibinfo  {journal}
  {Physica Scripta}\ }\textbf {\bibinfo {volume} {T76}},\ \bibinfo {pages}
  {165} (\bibinfo {year} {1998})}\BibitemShut {NoStop}%
\bibitem [{\citenamefont {Nakamura}\ \emph {et~al.}(1999)\citenamefont
  {Nakamura}, \citenamefont {Pashkin},\ and\ \citenamefont {Tsai}}]{NaPaTs99}%
  \BibitemOpen
  \bibfield  {author} {\bibinfo {author} {\bibfnamefont {Y.}~\bibnamefont
  {Nakamura}}, \bibinfo {author} {\bibfnamefont {Y.~A.}\ \bibnamefont
  {Pashkin}}, \ and\ \bibinfo {author} {\bibfnamefont {J.-S.}\ \bibnamefont
  {Tsai}},\ }\href {\doibase 10.1038/19718} {\bibfield  {journal} {\bibinfo
  {journal} {Nature}\ }\textbf {\bibinfo {volume} {398}},\ \bibinfo {pages}
  {786} (\bibinfo {year} {1999})},\ \Eprint
  {http://arxiv.org/abs/cond-mat/9904003} {arXiv:cond-mat/9904003
  [cond-mat.mes-hall]} \BibitemShut {NoStop}%
\bibitem [{\citenamefont {Nakamura}(2009)}]{Nak09}%
  \BibitemOpen
  \bibfield  {author} {\bibinfo {author} {\bibfnamefont {Y.}~\bibnamefont
  {Nakamura}},\ }\href {\doibase 10.1038/459516a} {\bibfield  {journal}
  {\bibinfo  {journal} {Nature}\ }\textbf {\bibinfo {volume} {459}},\ \bibinfo
  {pages} {516} (\bibinfo {year} {2009})}\BibitemShut {NoStop}%
\bibitem [{\citenamefont {Evans}\ and\ \citenamefont {Searles}(1994)}]{EvSe94}%
  \BibitemOpen
  \bibfield  {author} {\bibinfo {author} {\bibfnamefont {D.~J.}\ \bibnamefont
  {Evans}}\ and\ \bibinfo {author} {\bibfnamefont {D.~J.}\ \bibnamefont
  {Searles}},\ }\href {\doibase 10.1103/PhysRevE.50.1645} {\bibfield  {journal}
  {\bibinfo  {journal} {Physical Review E}\ }\textbf {\bibinfo {volume} {50}},\
  \bibinfo {pages} {1645} (\bibinfo {year} {1994})}\BibitemShut {NoStop}%
\bibitem [{\citenamefont {Gallavotti}\ and\ \citenamefont
  {Cohen}(1995)}]{GaCo95}%
  \BibitemOpen
  \bibfield  {author} {\bibinfo {author} {\bibfnamefont {G.}~\bibnamefont
  {Gallavotti}}\ and\ \bibinfo {author} {\bibfnamefont {E.~G.~D.}\ \bibnamefont
  {Cohen}},\ }\href {\doibase 10.1103/PhysRevLett.74.2694} {\bibfield
  {journal} {\bibinfo  {journal} {Physical Review Letters}\ }\textbf {\bibinfo
  {volume} {74}},\ \bibinfo {pages} {2694} (\bibinfo {year} {1995})},\ \Eprint
  {http://arxiv.org/abs/chao-dyn/9410007} {arXiv:chao-dyn/9410007 [nlin.CD]}
  \BibitemShut {NoStop}%
\bibitem [{\citenamefont {Jarzynski}(1997)}]{Jar97}%
  \BibitemOpen
  \bibfield  {author} {\bibinfo {author} {\bibfnamefont {C.}~\bibnamefont
  {Jarzynski}},\ }\href {\doibase 10.1103/PhysRevLett.78.2690} {\bibfield
  {journal} {\bibinfo  {journal} {Physical Review Letters}\ }\textbf {\bibinfo
  {volume} {78}},\ \bibinfo {pages} {2690} (\bibinfo {year} {1997})},\ \Eprint
  {http://arxiv.org/abs/cond-mat/9610209} {arXiv:cond-mat/9610209
  [cond-mat.stat-mech]} \BibitemShut {NoStop}%
\bibitem [{\citenamefont {Crooks}(1997)}]{Cro97}%
  \BibitemOpen
  \bibfield  {author} {\bibinfo {author} {\bibfnamefont {G.~E.}\ \bibnamefont
  {Crooks}},\ }\href {\doibase 10.1023/A:1023208217925} {\bibfield  {journal}
  {\bibinfo  {journal} {Journal of Statistical Physics}\ }\textbf {\bibinfo
  {volume} {90}},\ \bibinfo {pages} {1481} (\bibinfo {year}
  {1997})}\BibitemShut {NoStop}%
\bibitem [{\citenamefont {Kurchan}(1998)}]{Kur98}%
  \BibitemOpen
  \bibfield  {author} {\bibinfo {author} {\bibfnamefont {J.}~\bibnamefont
  {Kurchan}},\ }\href {\doibase 10.1088/0305-4470/31/16/003} {\bibfield
  {journal} {\bibinfo  {journal} {Journal of Physics A: Mathematical and
  General}\ }\textbf {\bibinfo {volume} {31}},\ \bibinfo {pages} {3719}
  (\bibinfo {year} {1998})},\ \Eprint {http://arxiv.org/abs/cond-mat/9709304}
  {arXiv:cond-mat/9709304 [cond-mat.stat-mech]} \BibitemShut {NoStop}%
\bibitem [{\citenamefont {Lebowitz}\ and\ \citenamefont
  {Spohn}(1999)}]{LeSp99}%
  \BibitemOpen
  \bibfield  {author} {\bibinfo {author} {\bibfnamefont {J.~L.}\ \bibnamefont
  {Lebowitz}}\ and\ \bibinfo {author} {\bibfnamefont {H.}~\bibnamefont
  {Spohn}},\ }\href {\doibase 10.1023/A:1004589714161} {\bibfield  {journal}
  {\bibinfo  {journal} {Journal of Statistical Physics}\ }\textbf {\bibinfo
  {volume} {95}},\ \bibinfo {pages} {333} (\bibinfo {year} {1999})},\ \Eprint
  {http://arxiv.org/abs/cond-mat/9811220} {arXiv:cond-mat/9811220
  [cond-mat.stat-mech]} \BibitemShut {NoStop}%
\bibitem [{\citenamefont {Jiang}\ \emph {et~al.}(2004)\citenamefont {Jiang},
  \citenamefont {Qian},\ and\ \citenamefont {Qian}}]{JiangQianQian}%
  \BibitemOpen
  \bibfield  {author} {\bibinfo {author} {\bibfnamefont {D.-Q.}\ \bibnamefont
  {Jiang}}, \bibinfo {author} {\bibfnamefont {M.}~\bibnamefont {Qian}}, \ and\
  \bibinfo {author} {\bibfnamefont {M.-P.}\ \bibnamefont {Qian}},\ }\href
  {\doibase 10.1007/b94615} {\emph {\bibinfo {title} {{Mathematical Theory of
  Nonequilibrium Steady States}}}},\ \bibinfo {series} {Lecture Notes in
  Mathematics}, Vol.\ \bibinfo {volume} {1833}\ (\bibinfo  {publisher}
  {Springer},\ \bibinfo {year} {2004})\ p.\ \bibinfo {pages} {276}\BibitemShut
  {NoStop}%
\bibitem [{\citenamefont {Ch{\'e}trite}\ and\ \citenamefont
  {Gaw\c{e}dzki}(2008)}]{ChGa08}%
  \BibitemOpen
  \bibfield  {author} {\bibinfo {author} {\bibfnamefont {R.}~\bibnamefont
  {Ch{\'e}trite}}\ and\ \bibinfo {author} {\bibfnamefont {K.}~\bibnamefont
  {Gaw\c{e}dzki}},\ }\href {\doibase 10.1007/s00220-008-0502-9} {\bibfield
  {journal} {\bibinfo  {journal} {Communications in Mathematical Physics}\
  }\textbf {\bibinfo {volume} {282}},\ \bibinfo {pages} {469} (\bibinfo {year}
  {2008})},\ \Eprint {http://arxiv.org/abs/0707.2725} {arXiv:0707.2725
  [math-ph]} \BibitemShut {NoStop}%
\bibitem [{\citenamefont {Sekimoto}(2010)}]{Sekimoto}%
  \BibitemOpen
  \bibfield  {author} {\bibinfo {author} {\bibfnamefont {K.}~\bibnamefont
  {Sekimoto}},\ }\href {\doibase 10.1007/978-3-642-05411-2} {\emph {\bibinfo
  {title} {{Stochastic Energetics}}}},\ \bibinfo {series} {Lecture Notes in
  Physics}, Vol.\ \bibinfo {volume} {799}\ (\bibinfo  {publisher} {Springer},\
  \bibinfo {year} {2010})\ p.\ \bibinfo {pages} {322}\BibitemShut {NoStop}%
\bibitem [{\citenamefont {Kurchan}(2000)}]{Kur00}%
  \BibitemOpen
  \bibfield  {author} {\bibinfo {author} {\bibfnamefont {J.}~\bibnamefont
  {Kurchan}},\ }\href@noop {} {\enquote {\bibinfo {title} {{A Quantum
  Fluctuation Theorem}},}\ } (\bibinfo {year} {2000}),\ \bibinfo {note}
  {preprint},\ \Eprint {http://arxiv.org/abs/cond-mat/0007360}
  {arXiv:cond-mat/0007360 [cond-mat.stat-mech]} \BibitemShut {NoStop}%
\bibitem [{\citenamefont {Mukamel}(2003)}]{Mu03}%
  \BibitemOpen
  \bibfield  {author} {\bibinfo {author} {\bibfnamefont {S.}~\bibnamefont
  {Mukamel}},\ }\href {\doibase 10.1103/PhysRevLett.90.170604} {\bibfield
  {journal} {\bibinfo  {journal} {Physical Review Letters}\ }\textbf {\bibinfo
  {volume} {90}},\ \bibinfo {pages} {170604} (\bibinfo {year} {2003})},\
  \Eprint {http://arxiv.org/abs/cond-mat/0302190} {arXiv:cond-mat/0302190
  [cond-mat.stat-mech]} \BibitemShut {NoStop}%
\bibitem [{\citenamefont {De~Roeck}\ and\ \citenamefont
  {Maes}(2004)}]{DeRoMa04}%
  \BibitemOpen
  \bibfield  {author} {\bibinfo {author} {\bibfnamefont {W.}~\bibnamefont
  {De~Roeck}}\ and\ \bibinfo {author} {\bibfnamefont {C.}~\bibnamefont
  {Maes}},\ }\href {\doibase 10.1103/PhysRevE.69.026115} {\bibfield  {journal}
  {\bibinfo  {journal} {Physical Review E}\ }\textbf {\bibinfo {volume} {69}},\
  \bibinfo {pages} {026115} (\bibinfo {year} {2004})},\ \Eprint
  {http://arxiv.org/abs/cond-mat/0309498} {arXiv:cond-mat/0309498 [cond-mat]}
  \BibitemShut {NoStop}%
\bibitem [{\citenamefont {Jarzynski}\ and\ \citenamefont
  {W\'ojcik}(2004)}]{JaWo04}%
  \BibitemOpen
  \bibfield  {author} {\bibinfo {author} {\bibfnamefont {C.}~\bibnamefont
  {Jarzynski}}\ and\ \bibinfo {author} {\bibfnamefont {D.~K.}\ \bibnamefont
  {W\'ojcik}},\ }\href {\doibase 10.1103/PhysRevLett.92.230602} {\bibfield
  {journal} {\bibinfo  {journal} {Physical Review Letters}\ }\textbf {\bibinfo
  {volume} {92}},\ \bibinfo {pages} {230602} (\bibinfo {year} {2004})},\
  \Eprint {http://arxiv.org/abs/cond-mat/0404475} {arXiv:cond-mat/0404475
  [cond-mat.stat-mech]} \BibitemShut {NoStop}%
\bibitem [{\citenamefont {Attal}\ and\ \citenamefont
  {Gaw\c{e}dzki}(2010)}]{AtGa09}%
  \BibitemOpen
  \bibfield  {author} {\bibinfo {author} {\bibfnamefont {S.}~\bibnamefont
  {Attal}}\ and\ \bibinfo {author} {\bibfnamefont {K.}~\bibnamefont
  {Gaw\c{e}dzki}},\ }\href@noop {} {\enquote {\bibinfo {title} {{Jarzynski
  equality for quantum dissipative dynamics}},}\ } (\bibinfo {year} {2010}),\
  \bibinfo {note} {{ENS notes}}\BibitemShut {NoStop}%
\bibitem [{\citenamefont {Campisi}\ \emph {et~al.}(2011)\citenamefont
  {Campisi}, \citenamefont {H\"anggi},\ and\ \citenamefont
  {Talkner}}]{CaHaTa11}%
  \BibitemOpen
  \bibfield  {author} {\bibinfo {author} {\bibfnamefont {M.}~\bibnamefont
  {Campisi}}, \bibinfo {author} {\bibfnamefont {P.}~\bibnamefont {H\"anggi}}, \
  and\ \bibinfo {author} {\bibfnamefont {P.}~\bibnamefont {Talkner}},\ }\href
  {\doibase 10.1103/revmodphys.83.771} {\bibfield  {journal} {\bibinfo
  {journal} {Review of Modern Physics}\ }\textbf {\bibinfo {volume} {83}},\
  \bibinfo {pages} {771} (\bibinfo {year} {2011})},\ \Eprint
  {http://arxiv.org/abs/1012.2268} {arXiv:1012.2268 [cond-mat.stat-mech]}
  \BibitemShut {NoStop}%
\bibitem [{\citenamefont {Ch{\'e}trite}\ and\ \citenamefont
  {Mallick}(2012)}]{ChMa12}%
  \BibitemOpen
  \bibfield  {author} {\bibinfo {author} {\bibfnamefont {R.}~\bibnamefont
  {Ch{\'e}trite}}\ and\ \bibinfo {author} {\bibfnamefont {K.}~\bibnamefont
  {Mallick}},\ }\href {\doibase 10.1007/s10955-012-0557-z} {\bibfield
  {journal} {\bibinfo  {journal} {Journal of Statistical Physics}\ }\textbf
  {\bibinfo {volume} {148}},\ \bibinfo {pages} {480} (\bibinfo {year}
  {2012})},\ \Eprint {http://arxiv.org/abs/1002.0950} {arXiv:1002.0950
  [cond-mat.stat-mech]} \BibitemShut {NoStop}%
\bibitem [{\citenamefont {Horowitz}\ and\ \citenamefont
  {Parrondo}(2013)}]{HoPa13}%
  \BibitemOpen
  \bibfield  {author} {\bibinfo {author} {\bibfnamefont {J.~M.}\ \bibnamefont
  {Horowitz}}\ and\ \bibinfo {author} {\bibfnamefont {J.~M.~R.}\ \bibnamefont
  {Parrondo}},\ }\href {\doibase 10.1088/1367-2630/15/8/085028} {\bibfield
  {journal} {\bibinfo  {journal} {New Journal of Physics}\ }\textbf {\bibinfo
  {volume} {15}},\ \bibinfo {pages} {085028} (\bibinfo {year} {2013})},\
  \Eprint {http://arxiv.org/abs/1305.6793} {arXiv:1305.6793
  [cond-mat.stat-mech]} \BibitemShut {NoStop}%
\bibitem [{\citenamefont {Albash}\ \emph {et~al.}(2013)\citenamefont {Albash},
  \citenamefont {Lidar}, \citenamefont {Marvian},\ and\ \citenamefont
  {Zanardi}}]{AlLiMaZa13}%
  \BibitemOpen
  \bibfield  {author} {\bibinfo {author} {\bibfnamefont {T.}~\bibnamefont
  {Albash}}, \bibinfo {author} {\bibfnamefont {D.~A.}\ \bibnamefont {Lidar}},
  \bibinfo {author} {\bibfnamefont {M.}~\bibnamefont {Marvian}}, \ and\
  \bibinfo {author} {\bibfnamefont {P.}~\bibnamefont {Zanardi}},\ }\href
  {\doibase 10.1103/physreve.88.032146} {\bibfield  {journal} {\bibinfo
  {journal} {Physical Review E}\ }\textbf {\bibinfo {volume} {88}},\ \bibinfo
  {pages} {032146} (\bibinfo {year} {2013})},\ \Eprint
  {http://arxiv.org/abs/1212.6589} {arXiv:1212.6589 [quant-ph]} \BibitemShut
  {NoStop}%
\bibitem [{\citenamefont {Solinas}\ \emph {et~al.}(2013)\citenamefont
  {Solinas}, \citenamefont {Averin},\ and\ \citenamefont {Pekola}}]{SoAvPe13}%
  \BibitemOpen
  \bibfield  {author} {\bibinfo {author} {\bibfnamefont {P.}~\bibnamefont
  {Solinas}}, \bibinfo {author} {\bibfnamefont {D.~V.}\ \bibnamefont {Averin}},
  \ and\ \bibinfo {author} {\bibfnamefont {J.~P.}\ \bibnamefont {Pekola}},\
  }\href {\doibase 10.1103/physrevb.87.060508} {\bibfield  {journal} {\bibinfo
  {journal} {Physical Review B}\ }\textbf {\bibinfo {volume} {87}},\ \bibinfo
  {pages} {060508(R)} (\bibinfo {year} {2013})},\ \Eprint
  {http://arxiv.org/abs/1206.5699} {arXiv:1206.5699 [quant-ph]} \BibitemShut
  {NoStop}%
\bibitem [{\citenamefont {Pekola}\ \emph {et~al.}(2016)\citenamefont {Pekola},
  \citenamefont {Suomela},\ and\ \citenamefont {Galperin}}]{PeSuGa16}%
  \BibitemOpen
  \bibfield  {author} {\bibinfo {author} {\bibfnamefont {J.~P.}\ \bibnamefont
  {Pekola}}, \bibinfo {author} {\bibfnamefont {S.}~\bibnamefont {Suomela}}, \
  and\ \bibinfo {author} {\bibfnamefont {Y.~M.}\ \bibnamefont {Galperin}},\
  }\href {\doibase 10.1007/s10909-016-1618-5} {\bibfield  {journal} {\bibinfo
  {journal} {Journal of Low Temperature Physics}\ ,\ \bibinfo {pages} {1}}
  (\bibinfo {year} {2016})},\ \Eprint {http://arxiv.org/abs/1602.00474}
  {arXiv:1602.00474 [cond-mat.mes-hall]} \BibitemShut {NoStop}%
\bibitem [{\citenamefont {Breuer}\ and\ \citenamefont
  {Petruccione}(2002)}]{BreuerPetruccione}%
  \BibitemOpen
  \bibfield  {author} {\bibinfo {author} {\bibfnamefont {H.-P.}\ \bibnamefont
  {Breuer}}\ and\ \bibinfo {author} {\bibfnamefont {F.}~\bibnamefont
  {Petruccione}},\ }\href {\doibase 10.1093/acprof:oso/9780199213900.001.0001}
  {\emph {\bibinfo {title} {{The theory of open quantum systems}}}},\ \bibinfo
  {edition} {reprint}\ ed.\ (\bibinfo  {publisher} {Oxford University Press},\
  \bibinfo {year} {2002})\ p.\ \bibinfo {pages} {625}\BibitemShut {NoStop}%
\bibitem [{\citenamefont {Wiseman}\ and\ \citenamefont
  {Milburn}(2009)}]{WiMi09}%
  \BibitemOpen
  \bibfield  {author} {\bibinfo {author} {\bibfnamefont {H.~M.}\ \bibnamefont
  {Wiseman}}\ and\ \bibinfo {author} {\bibfnamefont {G.~J.}\ \bibnamefont
  {Milburn}},\ }\href {www.cambridge.org/9780521804424} {\emph {\bibinfo
  {title} {{Quantum Measurement and Control}}}},\ \bibinfo {edition} {1st}\
  ed.\ (\bibinfo  {publisher} {Cambridge University Press},\ \bibinfo {year}
  {2009})\ p.\ \bibinfo {pages} {478}\BibitemShut {NoStop}%
\bibitem [{\citenamefont {van~den Berg}\ \emph {et~al.}(2015)\citenamefont
  {van~den Berg}, \citenamefont {Brange},\ and\ \citenamefont
  {Samuelsson}}]{BeBrSa15}%
  \BibitemOpen
  \bibfield  {author} {\bibinfo {author} {\bibfnamefont {T.~L.}\ \bibnamefont
  {van~den Berg}}, \bibinfo {author} {\bibfnamefont {F.}~\bibnamefont
  {Brange}}, \ and\ \bibinfo {author} {\bibfnamefont {P.}~\bibnamefont
  {Samuelsson}},\ }\href {\doibase 10.1088/1367-2630/17/7/075012} {\bibfield
  {journal} {\bibinfo  {journal} {arXiv}\ } (\bibinfo {year} {2015}),\
  10.1088/1367-2630/17/7/075012},\ \Eprint {http://arxiv.org/abs/1506.05674}
  {1506.05674} \BibitemShut {NoStop}%
\bibitem [{\citenamefont {Suomela}\ \emph {et~al.}(2016)\citenamefont
  {Suomela}, \citenamefont {Kutvonen},\ and\ \citenamefont
  {Ala-Nissila}}]{SuKuAlNi16}%
  \BibitemOpen
  \bibfield  {author} {\bibinfo {author} {\bibfnamefont {S.}~\bibnamefont
  {Suomela}}, \bibinfo {author} {\bibfnamefont {A.}~\bibnamefont {Kutvonen}}, \
  and\ \bibinfo {author} {\bibfnamefont {T.}~\bibnamefont {Ala-Nissila}},\
  }\href@noop {} {\bibfield  {journal} {\bibinfo  {journal} {eprint}\ }
  (\bibinfo {year} {2016})},\ \Eprint {http://arxiv.org/abs/1601.05317}
  {arXiv:1601.05317 [quant-ph]} \BibitemShut {NoStop}%
\bibitem [{\citenamefont {Ashcroft}\ and\ \citenamefont
  {Mermin}(1976)}]{AsMe76}%
  \BibitemOpen
  \bibfield  {author} {\bibinfo {author} {\bibfnamefont {N.~W.}\ \bibnamefont
  {Ashcroft}}\ and\ \bibinfo {author} {\bibfnamefont {D.}~\bibnamefont
  {Mermin}},\ }\href@noop {} {\emph {\bibinfo {title} {{Solid State
  Physics}}}},\ \bibinfo {edition} {1st}\ ed.\ (\bibinfo  {publisher} {Saunders
  College Publishing, Philadelphia},\ \bibinfo {year} {1976})\ p.\ \bibinfo
  {pages} {848}\BibitemShut {NoStop}%
\bibitem [{\citenamefont {Breuer}(2007)}]{Bre07}%
  \BibitemOpen
  \bibfield  {author} {\bibinfo {author} {\bibfnamefont {H.-P.}\ \bibnamefont
  {Breuer}},\ }\href {\doibase 10.1103/physreva.75.022103} {\bibfield
  {journal} {\bibinfo  {journal} {Physical Review A}\ }\textbf {\bibinfo
  {volume} {75}},\ \bibinfo {pages} {022103} (\bibinfo {year} {2007})},\
  \Eprint {http://arxiv.org/abs/quant-ph/0611208} {arXiv:quant-ph/0611208
  [quant-ph]} \BibitemShut {NoStop}%
\bibitem [{\citenamefont {Pellegrini}(2014)}]{Pel14}%
  \BibitemOpen
  \bibfield  {author} {\bibinfo {author} {\bibfnamefont {C.}~\bibnamefont
  {Pellegrini}},\ }\href {\doibase 10.1007/s10955-013-0910-x} {\bibfield
  {journal} {\bibinfo  {journal} {Journal of Statistical Physics}\ }\textbf
  {\bibinfo {volume} {154}},\ \bibinfo {pages} {838} (\bibinfo {year}
  {2014})}\BibitemShut {NoStop}%
\bibitem [{\citenamefont {Breuer}(2003)}]{Bre03}%
  \BibitemOpen
  \bibfield  {author} {\bibinfo {author} {\bibfnamefont {H.-P.}\ \bibnamefont
  {Breuer}},\ }\href {\doibase 10.1103/physreva.68.032105} {\bibfield
  {journal} {\bibinfo  {journal} {Physical Review A}\ }\textbf {\bibinfo
  {volume} {68}},\ \bibinfo {pages} {032105} (\bibinfo {year} {2003})},\
  \Eprint {http://arxiv.org/abs/quant-ph/0306047} {arXiv:quant-ph/0306047
  [quant-ph]} \BibitemShut {NoStop}%
\bibitem [{\citenamefont {Pusz}\ and\ \citenamefont
  {Woronowicz}(1978)}]{PuWo78}%
  \BibitemOpen
  \bibfield  {author} {\bibinfo {author} {\bibfnamefont {W.}~\bibnamefont
  {Pusz}}\ and\ \bibinfo {author} {\bibfnamefont {S.~L.}\ \bibnamefont
  {Woronowicz}},\ }\href {\doibase 10.1007/bf01614224} {\bibfield  {journal}
  {\bibinfo  {journal} {Communications in Mathematical Physics}\ }\textbf
  {\bibinfo {volume} {58}},\ \bibinfo {pages} {273} (\bibinfo {year}
  {1978})}\BibitemShut {NoStop}%
\bibitem [{\citenamefont {Alicki}(1979)}]{Ali79}%
  \BibitemOpen
  \bibfield  {author} {\bibinfo {author} {\bibfnamefont {R.}~\bibnamefont
  {Alicki}},\ }\href {\doibase 10.1088/0305-4470/12/5/007} {\bibfield
  {journal} {\bibinfo  {journal} {Journal of Physics A: Mathematical and
  General}\ }\textbf {\bibinfo {volume} {12}},\ \bibinfo {pages} {L103–L107}
  (\bibinfo {year} {1979})}\BibitemShut {NoStop}%
\bibitem [{\citenamefont {Wellstood}\ \emph {et~al.}(1994)\citenamefont
  {Wellstood}, \citenamefont {Urbina},\ and\ \citenamefont
  {Clarke}}]{WeUrCl94}%
  \BibitemOpen
  \bibfield  {author} {\bibinfo {author} {\bibfnamefont {F.~C.}\ \bibnamefont
  {Wellstood}}, \bibinfo {author} {\bibfnamefont {C.}~\bibnamefont {Urbina}}, \
  and\ \bibinfo {author} {\bibfnamefont {J.}~\bibnamefont {Clarke}},\ }\href
  {\doibase 10.1103/physrevb.49.5942} {\bibfield  {journal} {\bibinfo
  {journal} {Physical Review B}\ }\textbf {\bibinfo {volume} {49}},\ \bibinfo
  {pages} {5942} (\bibinfo {year} {1994})}\BibitemShut {NoStop}%
\bibitem [{\citenamefont {Cojocaru}\ and\ \citenamefont
  {Anghel}(2016)}]{CoAn16}%
  \BibitemOpen
  \bibfield  {author} {\bibinfo {author} {\bibfnamefont {S.}~\bibnamefont
  {Cojocaru}}\ and\ \bibinfo {author} {\bibfnamefont {D.-V.}\ \bibnamefont
  {Anghel}},\ }\href {\doibase 10.1103/physrevb.93.115405} {\bibfield
  {journal} {\bibinfo  {journal} {Physical Review B}\ }\textbf {\bibinfo
  {volume} {93}},\ \bibinfo {pages} {115405} (\bibinfo {year} {2016})},\
  \Eprint {http://arxiv.org/abs/1603.09061} {arXiv:1603.09061
  [cond-mat.mes-hall]} \BibitemShut {NoStop}%
\bibitem [{\citenamefont {Donvil}(2016)}]{Don16}%
  \BibitemOpen
  \bibfield  {author} {\bibinfo {author} {\bibfnamefont {B.}~\bibnamefont
  {Donvil}},\ }\emph {\bibinfo {title} {{Modeling and Simulation of a two-level
  system interacting with a calorimeter}}},\ \href@noop {} {Master's thesis},\
  \bibinfo  {school} {KU Leuven} (\bibinfo {year} {2016})\BibitemShut {NoStop}%
\bibitem [{\citenamefont {Suomela}\ \emph {et~al.}(2014)\citenamefont
  {Suomela}, \citenamefont {Solinas}, \citenamefont {Pekola}, \citenamefont
  {Ankerhold},\ and\ \citenamefont {Ala-Nissila}}]{SuSaPeAnAl14}%
  \BibitemOpen
  \bibfield  {author} {\bibinfo {author} {\bibfnamefont {S.}~\bibnamefont
  {Suomela}}, \bibinfo {author} {\bibfnamefont {P.}~\bibnamefont {Solinas}},
  \bibinfo {author} {\bibfnamefont {J.~P.}\ \bibnamefont {Pekola}}, \bibinfo
  {author} {\bibfnamefont {J.}~\bibnamefont {Ankerhold}}, \ and\ \bibinfo
  {author} {\bibfnamefont {T.}~\bibnamefont {Ala-Nissila}},\ }\href {\doibase
  10.1103/physrevb.90.094304} {\bibfield  {journal} {\bibinfo  {journal}
  {Physical Review B}\ }\textbf {\bibinfo {volume} {90}},\ \bibinfo {pages}
  {094304} (\bibinfo {year} {2014})},\ \Eprint {http://arxiv.org/abs/1404.0610}
  {arXiv:1404.0610 [quant-ph]} \BibitemShut {NoStop}%
\bibitem [{\citenamefont {van Hove}(1954)}]{vHo54}%
  \BibitemOpen
  \bibfield  {author} {\bibinfo {author} {\bibfnamefont {L.}~\bibnamefont {van
  Hove}},\ }\href {\doibase 10.1016/s0031-8914(54)92646-4} {\bibfield
  {journal} {\bibinfo  {journal} {Physica}\ }\textbf {\bibinfo {volume} {21}},\
  \bibinfo {pages} {517–54} (\bibinfo {year} {1954})}\BibitemShut {NoStop}%
\bibitem [{\citenamefont {Davies}(1974)}]{Dav74}%
  \BibitemOpen
  \bibfield  {author} {\bibinfo {author} {\bibfnamefont {E.~B.}\ \bibnamefont
  {Davies}},\ }\href {\doibase 10.1007/bf01608389} {\bibfield  {journal}
  {\bibinfo  {journal} {Communications in Mathematical Physics}\ }\textbf
  {\bibinfo {volume} {39}},\ \bibinfo {pages} {91} (\bibinfo {year}
  {1974})}\BibitemShut {NoStop}%
\bibitem [{\citenamefont {Davies}(1976)}]{Dav76}%
  \BibitemOpen
  \bibfield  {author} {\bibinfo {author} {\bibfnamefont {E.~B.}\ \bibnamefont
  {Davies}},\ }\href@noop {} {\emph {\bibinfo {title} {{Quantum Theory of Open
  Systems}}}}\ (\bibinfo  {publisher} {Academic Press},\ \bibinfo {year}
  {1976})\ p.\ \bibinfo {pages} {171}\BibitemShut {NoStop}%
\end{thebibliography}%

\end{document}